\documentclass[prd,aps]{revtex4}
\usepackage{graphicx}
\def \a {\alpha}
\def \b {\beta}
\def \g {\gamma}

\def \d {\delta}

\def \ep {\epsilon}

\def \f {\phi}
\def \ffi {\varphi}

\def \k {\kappa}
\def \l {\lambda}
\def \m {\mu}
\def \n {\nu}

\def \p {\pi}
\def \r {\rho}
\def \s {\sigma}
\def \t {\tau}

\def \o {\omega}

\def \der {\partial }
\def \nn {\nonumber}
\def \rov {\equiv}

\def \pul {{{\scriptstyle{\frac{1}{2}}}}}

\def \mm {\mbox{\quad }}

\def \msip {\rightarrow}

\def \BE {\begin{equation}}
\def \EE {\end{equation}}
\def \BDM {\begin{displaymath}}
\def \EDM {\end{displaymath}}
\def \BEAH {\begin{eqnarray*}}
\def \EEAH {\end{eqnarray*}}
\def \BEA {\begin{eqnarray}}
\def \EEA {\end{eqnarray}}
\def \BM {\begin{math}}
\def \EM {\end{math}}
\def \mm {\mbox{\quad }}

\def \msip {\rightarrow}

\def \pul {{{\scriptstyle{\frac{1}{2}}}}}

\begin{document}

\title{On the spinning C-metric \footnote{Published in 
{\it Gravitation: Following the Prague Inspiration} (Selected essays in honour of J. {Bi\v c\' ak}), Eds.
O. Semer\'ak, J. Podolsk\'y, and M. \v Zofka, World Scientific, Singapore (2002).}
}

\author{V. Pravda, A. Pravdov\'a}

\address{Mathematical Institute, 
Academy of Sciences, {\v Zitn\' a} 25,
115 67 Prague 1, Czech Republic \\
E-mail: pravda@math.cas.cz, pravdova@math.cas.cz}

\begin{abstract}
Physical interpretation of some
stationary and non-stationary regions of the~spinning
C-metric is presented. They represent different spacetime
regions  of a uniformly accelerated Kerr black hole.
Stability of geodesics corresponding to equilibrium points 
in a general stationary spacetime with an additional symmetry
is also studied and results are then applied to the~spinning
C-metric. 
\end{abstract}

\maketitle

\section{Introduction}

The~C-metric is a well known exact Petrov type D vacuum solution
of Einstein's field equations 
found in the~second decade of the~last century\cite{levi,weyl}. 
However, it was much later when its physical interpretation was 
found\cite{KinnWalker,BonnorN}. Its most physical region represents
a pair of Schwarzschild black holes
connected by a conical singularity and uniformly accelerated in opposite
directions along the~axis of axial symmetry.

The~C-metric is an example of boost-rotation symmetric 
spacetimes\cite{BicSchPRD,JibiEhlers,AVrev} corresponding 
to gravitational field of uniformly
accelerated ``particles'' of various kinds. Thus for the~physical
interpretation it is worthwhile to express the~C-metric 
in coordinates adapted to the~boost and rotation 
symmetries\cite{BonnorN}.

Several generalizations of the~C-metric 
(e.g., for charged black holes\cite{KinnWalker,Cornish-el}, for an external field
present\cite{ernst} or recently for two or more black holes
accelerated in both directions along the~axis of symmetry\cite{Dowker}) 
were found. Probably the~most important of them is 
the~spinning C-metric (the~SC-metric) 
found by Pleba\' nski and Demia\' nski\cite{PlebDem},
which is also of the~Petrov type D.
In the~present paper,  we restrict ourselves to the~case with non-vanishing
mass, angular momentum, and acceleration and we set
electric and magnetic charges,
the~NUT parameter, and the~cosmological constant equal to zero.
Then the~SC-metric in coordinates adapted to its special 
algebraical structure reads
\BEA
{\rm d}s^2=\frac{1}{H'} 
 \Bigl[ \frac{W'}{P} {\rm d}p^2  + \frac{P}{W'} 
\left( {\rm d}\sigma + q^2 {\rm d} \tau' \right)^2  
+ \frac{W'}{Q} {\rm d}q^2  - \frac{Q}{W'}
 \left( {\rm d}\tau' - p^2 {\rm d} \sigma \right)^2 \Bigr]\ ,
 \label{Plebmetr}
\EEA
where
\BEA
P&=&\ \ \gamma'-{ \varepsilon'}\,{p}^{2}+2\,m'{p}^{3}-\gamma'{p}^{4}
 \label{eqP} \ , \nonumber \\
Q&=&-\gamma'+{ \varepsilon'}\,{q}^{2}+2\,m'{q}^{3}+\gamma'{q}^{4} 
\label{eqQ} \ ,\nn\\
H'&=&(p+q)^2\ ,\nn\\
W'&=&\ 1+(pq)^2\ \nn
\EEA
with $\gamma'$, $\varepsilon'$, and $m'$ being constant.

By performing certain limiting procedures\cite{PlebDem} 
 that remove the~acceleration or rotation 
parameters one obtains the~Kerr solution or the~C-metric, respectively.

Physical properties of the SC-metric were studied in 
\cite{FZ,bivoj,Letelier}.
Stationary regions
of the~SC-metric (\ref{Plebmetr}) and Killing horizons
were identified in \cite{bivoj}. Then the~most physical stationary region
was transformed into the~Weyl-Papapetrou coordinates and finally
to coordinates adapted to the~boost and rotation symmetries.
The~SC-metric is the~only  example of boost-rotation symmetric
spacetimes with spinning sources\cite{av-brs} known today. 

In the~present paper, we start with the~SC-metric in a better
chosen parametrization than in (\ref{Plebmetr}) that makes
the~physical interpretation more transparent.
As in \cite{bivoj}, we first transform stationary regions of the~SC-metric
to the~Weyl-Papapetrou form (Sec.~\ref{sec-Weyl}) and then 
to coordinates adapted  to the~boost and rotation symmetries
(Sec.~\ref{sec-brs}),
where non-stationary radiative regions appear. It is shown 
that these non-stationary regions 
(inside and outside the~null cone of the~origin)
are in fact already contained in the~original form
of the~SC-metric. This also enables us to locate 
${\cal J^+}$ and $I^+$.

It is shown in \cite{bivoj}  that the~most physical stationary 
region $\cal{B}$ corresponds to a field of a uniformly
accelerated Kerr black hole. Here, another stationary region is 
interpreted as a uniformly accelerated white hole with a ring 
singularity that in fact corresponds to the~uniformly accelerated
asymptotically flat
``interior'' of the~Kerr solution.

In Sec.~\ref{sec-geod}, geodesics and their stability are studied
at first generally for an arbitrary stationary metric with another
symmetry and then results are applied to
the~stationary region ${\cal B}$ of the~SC-metric.


\section{The~spinning C-metric in the \lowercase{$\{ \t$, $x$, $y$, $\f\}$} coordinates}

Since the~parameters $\gamma'$ and $\varepsilon'$ occurring 
in (\ref{Plebmetr})
do not have a straightforward connection with the~angular momentum
and acceleration of the~black hole,
we start with the~SC-metric 
with rescaled parameters and coordinates\cite{Letelier}
\BE
{\rm d}s^2=\frac{1}{H}\left[ \frac{W }{F} {\rm d}y^2 
+ \frac{W}{ G} {\rm d}x^2 +\frac{G}{W} ({\rm d} \f +aAy^2 {\rm d}\t)^2
- \frac{F}{W} ({\rm d} \t-aAx^2 {\rm d}\f)^2 \right]\ , \label{rotcmet}
\EE
where 
\BEA
F&=&-\delta-\varepsilon y^2-2mAy^3+(aA)^2\delta y^4 \ ,\nonumber \\
G&=&\delta+\varepsilon x^2-2mAx^3-(aA)^2\delta x^4\ ,\nonumber \\
H&=&A^2(x+y)^2\ ,\nn\\
W&=&1+(aAxy)^2\  \nonumber 
\EEA
with $m$, $a$, $A$, $\delta$, and $\varepsilon$ being constant.
The~metrics (\ref{Plebmetr}) and (\ref{rotcmet}) are related by the~trivial
transformation 
\BEA
p = \sqrt{aA}\ x\ ,\mm q = \sqrt{aA}\ y \ ,\mm   
\sigma =\sqrt{\frac{a}{A^3}}\ \f\ ,\mm 
\tau' = \sqrt{\frac{ a}{A^3}}\ \t\ ,\nn\\
\gamma' = A^2 \delta\ ,\mm \varepsilon' = -\frac{A \varepsilon}{a}\ ,\mm 
m' = -m\sqrt{\frac{A^3}{a^3}}\ .\mm\mm\mm\mm\nn
\EEA
Notice that for 
\BE
\delta = 1\ ,\mm   \varepsilon = -1\ ,\label{kinemat}
\EE 
and $a=0$ we get the~standard form of the non-spinning
C-metric. In the~following, we assign the~foregoing values (\ref{kinemat}) to
 the~kinematical parameters $\delta$ and $\varepsilon$.

In the~present paper, we study only the~case when the~polynomial
$F=-1+y^2-2mAy^3+(aA)^2y^4$
has four distinct real roots $y_1\ ,\dots,\ y_4$. This is satisfied if
the~conditions
\BE
m^2>\frac{8}{9}\ a^2 
\mm\mbox{and}\mm \a_1-\a_2 < A^2 < \a_1+\a_2\ ,\label{podm-rootb}
\EE
where
\BDM
\a_1=\frac{1}{32a^6} \left(36\,{a}^{2}{m}^{2}-8\,{a}^{4}-27\,{m}^{4}\right)
\ ,\mm
\a_2=\frac{m}{32a^6} \left(9\,{m}^{2}-8\,{a}^{2}\right)^{3/2}\ ,\nn
\EDM
hold. Notice that ${\a_1}^2-{\a_2}^2=(a^2-m^2)/(16a^2)$.
For $m^2>a^2$, $\a_1-\a_2$ is negative and then $A^2$ has to fulfil 
only the~upper constraint. 
In the~limit $ a \rightarrow 0$, the~upper constraint for $A^2$, $\a_1+\a_2$, turns out to be
$1/(27 m^2)$, which is the~same as for the~non-spinning C-metric.
If $y_i$ are the~roots of $F$ then $-y_i$ are the~roots of 
$G$ since $F(y)=-G(-y)$. 

\begin{figure}
\begin{center}
\includegraphics*{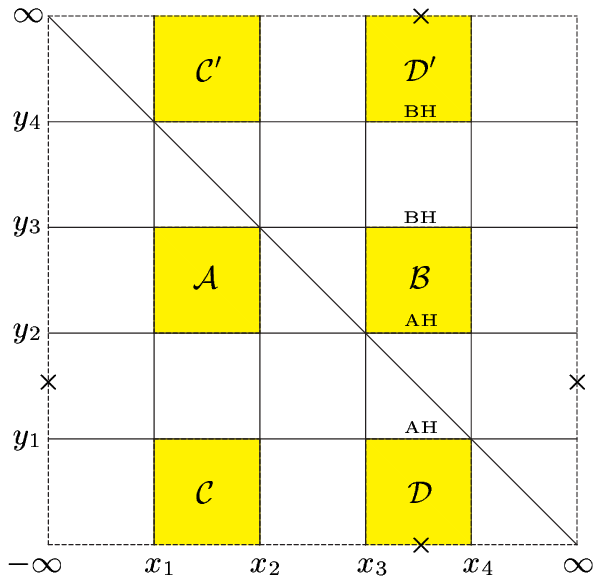}
\end{center}
\caption{The~structure of the~SC-metric (\ref{rotcmet})  
in the~coordinates
$\{ \t$, $x$, $y$, $\f\}$: The~metric has the~signature
$+2$ for $G>0$, i.e., in the~second and fourth columns from the~left,
and is stationary for $F>0$ in the~shaded squares. 
Curvature singularities at points $(x=0$, $y=\pm \infty)$ and 
$(x=\pm \infty$, $y=0)$ are denoted by crosses. The~black hole horizons
and  the~acceleration horizons are labeled by BH and AH, respectively.
The~line $x+y=0$, where curvature invariants (\ref{inva}) and (\ref{invb})
vanish, is also indicated.}
\label{ctverce}
\end{figure}

Similarly as in \cite{bivoj}, it can be shown that the~metric (\ref{rotcmet}) 
has the~signature $+2$ for $G>0$ and is then stationary for $F>0$.
In the~coordinates $\{ \t$, $x$, $y$, $\f\}$, there are four stationary 
regions (see Fig.~\ref{ctverce}) 
${\cal A}$, ${\cal B}$, ${\cal C}\cup {\cal C}'$, and 
${\cal D}\cup {\cal D}'$ (as in \cite{bivoj}, we identify $y=\infty$
and $y=-\infty$) and
each of them can be transformed to the~stationary 
Weyl-Papapetrou form (see Sec.~\ref{sec-Weyl}).

Curvature invariants
\BEA
R^{\a\b\g\d}R_{\a\b\g\d}&=&
              48m^2\frac{H^3}{W^6}(W-2)(W^2-16W+16)\ ,\label{inva}\\
\pul\ep^{\a\b\s\r}{R_{\s\r}}^{\m\n}R_{\a\b\m\n}&=&
             96aAm^2xy\frac{H^3}{W^6}(3W-4)(W-4)\ \label{invb}
\EEA
suggest that there are curvature singularities at points $(x=0$, $y=\pm \infty)$
and $(x=\pm \infty$, $y=0)$. The second curvature invariant also indicates that 
the~constant $a$ is proportional to the angular momentum of the~source \cite{Ciuf}. 
Both invariants vanish for $x+y=0$. Later it will be shown that ${\cal J}^+$,
$I^+$, and $I^0$ are located there (see Fig.~\ref{fig-nekonecno}).

Killing horizons, which are located at $y=y_i$, correspond
to the~black hole or acceleration horizons \cite{bivoj}  
(see Fig.~\ref{ctverce}).

\section{Transformation to the~Weyl-Papapetrou coordinates}
\label{sec-Weyl}

As was mentioned earlier, each of the four stationary regions 
${\cal A}$, ${\cal B}$, ${\cal C}\cup {\cal C}'$, and 
${\cal D}\cup {\cal D}'$ can be transformed
to the~stationary Weyl-Papapetrou coordinates 
$\{ {\bar t}$, ${\bar \r}$, ${\bar z}$, ${\bar \ffi}\}$
\BE
{\rm d}s^2={\rm e}^{-2U}
        [ {\rm e}^{2\n} ( {\rm d} {\bar \r}^2+{\rm d}{\bar z}^2)
      +{\bar \r}^2{\rm d}{\bar \ffi}^2] 
-{\rm e}^{2U}({\rm d}{\bar t}+{\bar \o}{\rm d}{\bar \ffi})^2\ , 
\label{dsWeyl}
\EE
where the~metric functions $U$, $\n$, and ${\bar \o}$ are functions of ${\bar \r}$
and ${\bar z}$. The~transformation 
\BEA
\bar{\rho}^2 &=& {\cal K}^2 FGH^{-2}\ ,  \label{transW1} \\
\bar z  &=& {\cal K} [1+mAxy(x-y)+xy-a^2A^2x^2y^2]H^{-1}\ , 
\label{transW2}\\
\t &=& \kappa_1 \bar t  + \kappa_2 \bar \ffi \ ,\\
\f &=& \kappa_3 \bar t + \kappa_4 \bar \ffi \ ,\\
{\cal K} &=& \kappa_2 \kappa_3 - \kappa_1 \kappa_4 \ ,\label{transW5}
\EEA
where $\k_1\dots\k_4$ are constant, can be  found
similarly as in \cite{bivoj}.
It turns out that it is convenient to choose 
\BE
{\cal K}=A \label{K}
\EE
since then the~SC-metric  in the~Weyl-Papapetrou coordinates
with appropriately chosen constants $\k_1\dots\k_4$ 
yields the~Kerr solution 
in the~limit $A\msip 0$. In the~following, we assume that (\ref{K})
holds.

In the~present paper, $\t$, $x$, $y$, $\f$, ${\bar \ffi}$, $\k_2$, and $\k_4$ 
are dimensionless and
${\bar t}$, ${\bar \r}$, ${\bar z}$, $m$, $a$, $1/A$,  $1/\k_1$,  and
$1/\k_3$ have the~dimension of length.   

In order to express the~SC-metric  in the~Weyl-Papapetrou form 
(\ref{dsWeyl}) we have to find the~inverse transformation to 
(\ref{transW1}), (\ref{transW2}).
Let us first define $R_i$
\BE
R_i\rov \sqrt{\bar \rho^2 + (\bar z-\bar z_i)^2} \ ,\mm i=1\dots 3\ ,
\label{eqRi}
\EE
where ${\bar z}_i$ are the~roots of the cubic equation
\BE
2A{\bar z}_i^3-{\bar z}_i^2+2Aa^2{\bar z}_i+m^2-a^2=0\ .\label{kubrce}
\EE
It can be shown that
\BDM
{R_i}^2=
\frac{ \left(\frac{ A_i}{A \bar z_i}  
+ \frac{A^2 m \bar z_i}{A_i}(x-y) -A_i xy  \right)^2}{A^2(x+y)^2}\ ,
\nn 
\EDM
where
\BDM
A_i^2=A^2(2 A a^2 \bar z_i  +m^2-a^2)\ .\nn
\EDM
Then
\BE
R_i = \epsilon_i \frac{ \left(\frac{ A_i}{A \bar z_i}  
+ \frac{A^2 m \bar z_i}{A_i}(x-y) -A_i xy
  \right)}{A(x+y)} \ ,
\label{eqRi2}
\EE
where $\epsilon_i = \pm 1$. Each of the~stationary regions is characterized
by a different combination of $\ep_i$:\vspace{2mm}
\begin{center}
\begin{tabular}{rccc}
${\cal A}:\mm$                 & $\ep_1=-1\ , $ & $\ep_2= -1\ ,$ & $\ep_3=+1\ , $ \\
${\cal B}:\mm$                 & $\ep_1=-1\ , $  & $\ep_2= +1\ , $ & $\ep_3=-1\ , $ \\
${\cal C} \cup {\cal C}':\mm$  & $\ep_1=+1\ , $  & $\ep_2= +1\ , $  & $\ep_3=+1\ , $  \\
${\cal D} \cup {\cal D}':\mm$ & $\ep_1=+1\ , $  & $\ep_2= -1\ , $ & $\ep_3= -1\ .$ \\
\end{tabular}
\end{center}
\vspace{2mm}

Now solving Eqs. (\ref{eqRi2}) for $x$ and $y$, 
the~inverse transformation can be found
\BE
x=\frac{F_0+F_1}{F_2} \ , \quad  y=\frac{F_0-F_1}{F_2}  \ ,\label{xy-rhoz}
\EE
where
\BEA
F_0 &=& \epsilon_1\epsilon_2\epsilon_3(m^2-a^2)A^2({\bar z}_1-{\bar z}_2)({\bar z}_2-{\bar z}_3)({\bar z}_3-{\bar z}_1)\ ,
 \nonumber \\
F_1 &=& 
  {\bar z}_1 A_2 A_3 \epsilon_3 \epsilon_2 ({\bar z}_3-{\bar z}_2) R_1
+ A_1 {\bar z}_2 A_3  \epsilon_1\epsilon_3 ({\bar z}_1-{\bar z}_3) R_2
+A_1 A_2 {\bar z}_3 \epsilon_2 \epsilon_1 ({\bar z}_2-{\bar z}_1) R_3 \ ,\nonumber \\
F_2 &=& (m^2-a^2) A \left[
\epsilon_2 \epsilon_3 ({\bar z}_2-{\bar z}_3) A_1  R_1
+\epsilon_3 \epsilon_1 ({\bar z}_3-{\bar z}_1) A_2  R_2
+\epsilon_1 \epsilon_2 ({\bar z}_1-{\bar z}_2) A_3 R_3 \right] \ .\nonumber 
\EEA

From the~transformation (\ref{transW1})--(\ref{transW5}) follows
\BEA
g_{{\bar \r}{\bar\r}}&=& \frac{W}{HF
       ({{\bar \r},_y}^2+{{\bar z},_y}^2)}\ , \nn\\ 
g_{\bar t \bar t} &=& \frac{-\left ({\it \kappa_1}
-a{x}^{2}{\it \kappa_3}\,A\right )^{2}F+\left ({\it \kappa_3}+a{y
}^{2}{\it \kappa_1}\,A\right )^{2}G}{H W}\ , \nn\\ 
g_{\bar \ffi \bar \ffi} &=& \frac{\left ({\it \kappa_2}
-a{x}^{2}{\it \kappa_4}\,A\right )^{2}F-\left ({\it \kappa_4}+a{y}
^{2}{\it \kappa_2}\,A\right )^{2}G}{H W}\ ,\label{grrgttfftf}\\
g_{\bar t \bar \ffi} &=& \frac{-\left ({\it \kappa_2}
-a{x}^{2}{\it \kappa_4}\,A\right )\left ({\it \kappa_1}-a{x}^{2}
{\it \kappa_3}\,A\right )F+\left ({\it \kappa_4}
+a{y}^{2}{\it \kappa_2}\,A\right )\left 
({\it \kappa_3}+a{y}^{2}{\it \kappa_1}\,A\right )G}{H W} \nn
\EEA
with $x$ and $y$ given by (\ref{xy-rhoz}).
The~Weyl-Papapetrou metric functions  are now
\BE
{\rm e}^{2U}=-g_{\bar t \bar t} \ ,\mm 
{\rm e}^{2\n}=-g_{{\bar \r}{\bar\r}}g_{\bar t \bar t}\ ,\mm
{\bar \o}=g_{\bar t \bar \ffi} /g_{\bar t \bar t} \ .\label{fceWeyl}
\EE
Substituting Eqs.~(\ref{xy-rhoz}) and (\ref{grrgttfftf}) into (\ref{fceWeyl}) we
obtain the~SC-metric in the~Weyl-Papapetrou
coordinates. Unfortunately, these expressions are very complicated 
and can  be handled only
by computer manipulations. In the non-spinning case, our approach gives again
long formulas, whereas Godfrey\cite{Godfrey} and Bonnor\cite{BonnorN},
after performing very long calculations, arrived at considerably simpler
equivalent results. 
A similar simplification may be probably also possible  in the~spinning case. 
The~spinning soliton generalization\cite{Letelsolit}  of the~non-spinning C-metric 
might be helpful in this task.

The~regularity condition on the~axis reads
\BDM
{\lim_{ {\bar \r}_0\msip 0} \frac{1}{2\p} 
        \frac{\int_0^{2\p} \sqrt {
          g_{{\bar \ffi}{\bar \ffi}}
       ({\bar \r}_0,{\bar z} )}
        {\rm d}{\bar \ffi}}
             {\int_0^{{\bar \r}_0}\sqrt{g_{{\bar \r}{\bar \r}}
                ({\bar \r},{\bar z} )}
         {\rm d}{\bar \r}}=1} \ ,\nn  
\EDM
from which follows
\BEA
\n&=&0\ ,\label{regn}\\
{\bar \o}&=&0\ \label{regomega}
\EEA
on the~axis. When both conditions  (\ref{regn})  and (\ref{regomega}) 
are satisfied then the~axis is regular. If only (\ref{regomega})  holds
then there is a conical singularity (a string or a strut);
if none of these conditions is satisfied a spinning string (conical and 
torsion\cite{letolstring,bonnor} singularity) is present
and, in its vicinity, a region with closed timelike curves occurs.
Vacuum Einstein's equations allow us to multiply ${\rm e}^{2\n}$ by a constant
$\k_5$. Later we will adjust this constant 
to regularize some parts of the~axis.

Let us now restrict ourselves to the~most physically
interesting stationary region ${\cal B}$ of the~SC-metric (\ref{rotcmet}),
which was studied in the~non-spinning and spinning cases in papers 
\cite{BonnorN,cornish,AVrev,Letelier} and \cite{bivoj,Letelier}, 
respectively. 

\begin{figure}
\begin{center}
\includegraphics*{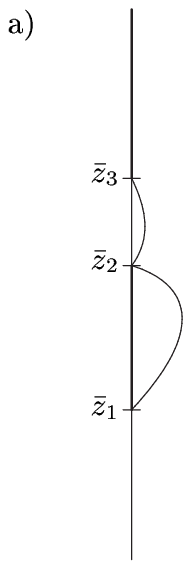} 
\hspace{3cm}
\includegraphics*{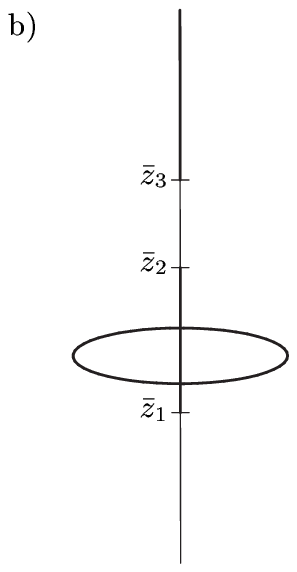} 
\end{center}
\caption{a) The~${\cal B}$  region of the~SC-metric in 
the~Weyl-Papapetrou coordinates: The~black hole and acceleration
horizons are located on the~axis ${\bar\r}=0$ at 
${\bar z}\in$ $({\bar z}_1,\ {\bar z}_2)$ and
at ${\bar z}\in$ $({\bar z}_3,\ \infty)$, respectively.
There is an ergoregion 
in the~vicinity of the~black hole horizon  
and a region with closed timelike curves
in the~neighbourhood of the~spinning string
between ${\bar z}_2$ and ${\bar z}_3$. \newline
b) The~${\cal D}\cup {\cal D}'$  region of the~SC-metric in 
the~Weyl-Papapetrou coordinates: The~white hole and acceleration
horizons are located on the~axis  at 
${\bar z}\in$ $({\bar z}_1,\ {\bar z}_2)$ and
 ${\bar z}\in$ $({\bar z}_3,\ \infty)$, respectively.
The~ring singularity with the~radius ${\bar\r}=a$
 appears at ${\bar z}=0$.}
\label{BaD-Weyl}
\end{figure}

In the~Weyl-Papapetrou form, the~black hole horizon is located
on the~axis between ${\bar z}_1$ and ${\bar z}_2$, the~acceleration
horizon extends from ${\bar z}_3$  to $\infty$, and there may occur
conical singularities or spinning strings on the~rest of the~axis
depending on the~values of the~constants $\kappa_1\dots\kappa_5$
(see Fig.~\ref{BaD-Weyl} a).

Let us now fix the~appropriate values of 
the~constants $\kappa_1\dots\kappa_5$:

Requiring  $g_{\bar t \bar \ffi}=0$ and $g_{\bar \ffi \bar \ffi} \not= 0$ 
on the~acceleration horizon (where $y=y_2$, $F=0$)
leads to the~condition
\BE
\kappa_3+ay_2^2A\kappa_1 = 0 \ . \label{rcekap3}
\EE
Demanding  further that there be no torsion singularity on the~axis for 
${\bar z}<\bar z_1$ (where $G=0$, $x=x_3=-y_2$),
i.e., ${\bar \o}$ and thus also $g_{\bar t \bar \ffi}$ vanish there
(\ref{regomega}), we arrive at
\BE
\kappa_2-ay_2^2A\kappa_4=0 \ .\label{rcekap2}
\EE
Equivalent conditions were obtained in \cite{bivoj} requiring 
the~appropriate asymptotical behaviour of the metric
in the~coordinates adapted to the~boost and rotation symmetries
discussed in Sec.~\ref{sec-brs}.  

From Eqs.~(\ref{K}), (\ref{rcekap3}), and (\ref{rcekap2}) we obtain
\BE
\k_4=-\frac{A}{\k_1(1+a^2A^2{y_2}^4)}\ .\label{rcekap4}
\EE

Further requirement on the~axis -- to be regular for ${\bar z}<\bar z_1$ 
(\ref{regn}) -- leads to
\BE
\k_5=-\frac{A^2(-1+4y_2(mA-a^2A^2y_2)+(a^2-m^2)A^2{y_2}^4)}
{{\k_1}^2(1+a^2A^2{y_2}^4)^2}\ .\label{rcekap5}
\EE
By Eqs.~(\ref{rcekap3})--(\ref{rcekap5}) together with
Eq.~(\ref{rcekap1}) the~constants $\k_1\dots\k_5$ are uniquely determined.

The~``angular velocity''  of the~black hole horizon $y=y_3$ 
turns out to be
\BE
\Omega_{H}\rov -\frac{g_{{\bar t}{\bar t}}}{g_{{\bar t}{\bar \ffi}}}_{|
y=y_3}=\frac{aA^2({y_3}^2-{y_2}^2)}{1+(aAy_2y_3)^2}\ ,\label{OmHp}
\EE
where Eqs.~(\ref{grrgttfftf}), (\ref{rcekap3})--(\ref{rcekap4}),
and (\ref{rcekap1}) were employed. Notice that the~expression
(\ref{OmHp}) differs from $\Omega_H$ of Letelier and Oliveira \cite{Letelier}
since we use a different coordinate system that  does not rotate 
asymptotically. 

For $a<m$ and the~acceleration $A\msip 0$, Eq.~(\ref{OmHp}) 
reduces to 
\BDM
\lim_{ A\msip 0} \Omega_{H}=\frac{a}{2m(m+\sqrt{m^2-a^2})}\ \nn
\EDM
that is equal to $\Omega_H$ for the~outer 
horizon of the~Kerr metric. 

For $a<m$ and $A<<1$, the~roots of (\ref{kubrce}) are
\BEA
{\bar z_1}&=& -\sqrt{m^2-a^2}+m^2 A+{\cal O}(A^2)\ ,\nn\\
{\bar z_2}&=&\sqrt{m^2-a^2}+m^2 A+{\cal O}(A^2)\ ,\nn\\
{\bar z_3}&=&\frac{1}{2A}-2m^2A+{\cal O}(A^2)\ .\nn
\EEA
In the~limit $A\msip 0$, there remains only the~black hole horizon 
on the~axis between
$-\sqrt{m^2-a^2}$ and $+\sqrt{m^2-a^2}$ since ${\bar z_3}\msip\infty$.
This indicates that in this limit the~SC-metric
in the~Weyl-Papapetrou coordinates with the~appropriate choice
of $\k_1\dots\k_5$ approaches the~Kerr solution. Indeed, the~metric functions
${\rm e}^{2U}$, ${\rm e}^{2\n}$, and ${\bar \o}$ with $\k_1\dots\k_5$
given by (\ref{rcekap3})--(\ref{rcekap5}) and (\ref{rcekap1})
go to the~corresponding metric functions of the~Kerr metric in 
the~Weyl-Papapetrou coordinates as given,  e.g., in \cite{kramer}.

In this section, we have transformed the~region ${\cal B}$
of the~SC-metric into the~Weyl-Papapetrou coordinates.
It is similarly possible to transform  the~region 
${\cal D}\cup {\cal D}'$ there, however, another
choice of the~constants $\k_1\dots\k_5$ would be appropriate
in Eqs.~(\ref{transW1})--(\ref{transW5}). 
The~white hole horizon
appears on the~axis between ${\bar z}_1$ and ${\bar z}_2$ 
and the~acceleration horizon for ${\bar z}>{\bar z}_3$.
Moreover there is a ring singularity at ${\bar z}=0$
and ${\bar \r}=a$ (corresponding to $x=0$ and $y=\pm \infty$), 
where curvature invariants go to infinity.
For the~limit $A\msip 0$ we obtain the~region $r\in (-\infty,\ r_-)$
of the~Kerr spacetime, where $r$ is the~Boyer-Lindquist coordinate.
The~region ${\cal D}\cup {\cal D}'$ thus corresponds to
the~spacetime of a uniformly accelerated white hole with a ring singularity.

\section{The~spinning C-metric in the~coordinates adapted 
to the~boost and rotation symmetries}
\label{sec-brs}

\begin{figure}
\begin{center}
\includegraphics*{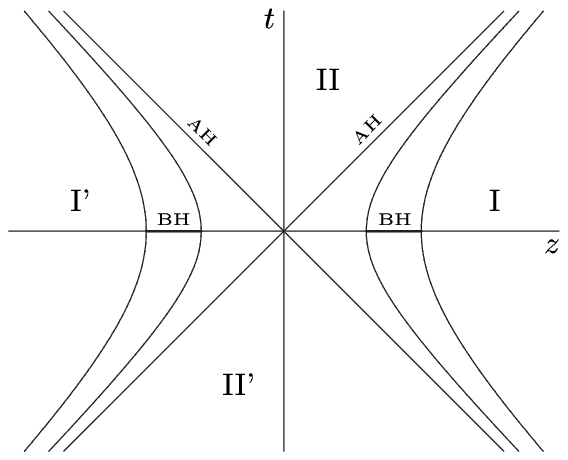} 
\end{center}
\caption{The~stationary region ${\cal B}$ of the~SC-metric 
transformed to the~coordinates adapted to the~boost
and rotation symmetries corresponds to the~region I under
the~roof, $z^2-t^2>0$, where the~roof is the~acceleration horizon,
denoted by AH. Under the~roof, there is also another identical stationary
region I' that corresponds to a second black hole
accelerated along the~symmetry axis in the~opposite direction. 
The~non-stationary, radiative regions II, II' appear above the~roof, 
$z^2-t^2<0$. This figure covers only the~outer part of the~spacetime
of a uniformly accelerated
Kerr black hole outside the~exterior black hole horizon (BH).
A similar picture, connected with ${\cal D}\cup {\cal D}'$,
would describe the~inner part of the~spacetime 
of a uniformly accelerated
Kerr black hole inside the~interior black hole horizon.}
\label{fig-boost}
\end{figure}

Now let us transform the~region ${\cal B}$ into the~coordinates
adapted to the~boost and rotation symmetries
$\{ t$, $\r$, $z$, $\ffi\}$. The~transformation \cite{BS,BonnorN} 
\BEA
{\bar \r}^2&=&A^2\r^2(z^2-t^2)\ ,\nn\\
{\bar \ffi}&=&\ffi\ ,\label{tr-Weyl-brs}\\
{\bar z}-{\bar z}_3&=&\pul A(\r^2+t^2-z^2)\ ,\nn\\
{\bar t}&=&\frac{1}{A}{\rm arctanh}\left(\frac{t}{z}\right)\ \nn
\EEA
leads to the~metric of boost-rotation symmetric spacetimes with
spinning sources\cite{bivoj,av-brs}
\BEA
{\rm d}s^2&=&{\rm e}^\l {\rm d}\r^2 +\r^2{\rm e}^{-\m}{\rm d}\ffi^2 
-2\o{\rm e}^\m (z{\rm d}t-t{\rm d}z){\rm d}\ffi
-\o^2{\rm e}^\m(z^2-t^2){\rm d}\ffi^2\ \nn\\
&+&
\frac{1}{z^2-t^2}[({\rm e}^\l z^2-{\rm e}^\m t^2) {\rm d}z^2
                  -2zt ({\rm e}^\l-{\rm e}^\m) {\rm d}z\ {\rm d}t
                  -({\rm e}^\m z^2-{\rm e}^\l t^2){\rm d} t^2],
\label{dsbrs}
\EEA
where 
$\m$, $\l$, and $\o$ -- functions of $\r^2$ and $z^2-t^2$ -- are given by 
\BEA
{\rm e}^\m&=&\frac{{\rm e}^{2U}}{A^2(z^2-t^2)}\ ,\nn\\
{\rm e}^\l&=&\k_5\frac{A^2{\rm e}^{2\n}}{{\rm e}^{2U}}(\r^2+z^2-t^2)\ ,
\label{em-eu}\\
\o&=&A{\bar \o}\ .\nn
\EEA
As a consequence of the~transformation (\ref{tr-Weyl-brs})
a second black hole accelerated along the~symmetry axis
in the~opposite direction appears (see Fig.~\ref{fig-boost}).

The~spacetime described by the~metric (\ref{dsbrs}) contains
stationary and non-stationary regions separated by a so-called ``roof''
given by $z^2-t^2=0$ (the~acceleration horizon); the~spacetime is
stationary ``bellow the~roof'' ($z^2-t^2>0$) and radiative ``above
the~roof'' ($z^2-t^2<0$), see Fig.~\ref{fig-boost}.
The~regularity condition of the~roof\cite{bivoj,av-brs}
\BE
\l(\r^2,0) =\m(\r^2,0)\ \nn
\EE
yields the~condition for $\k_1$
\BE
\k_1=\frac{A}{\sqrt{1+a^2A^2{y_2}^4}}\ .\label{rcekap1}
\EE

The~axis regularity condition\cite{bivoj,av-brs} 
\BE
\o(0,z^2-t^2) =0\ ,\mm \l(0,z^2-t^2) +\m(0,z^2-t^2) =0\ \nn
\EE
is satisfied on the~outer parts of the~axis 
thanks to Eqs.~(\ref{rcekap2}) and (\ref{rcekap5}).

The~metric functions (\ref{em-eu}) are in fact very complicated since
we have to use Eqs.~(\ref{eqRi}), (\ref{xy-rhoz})--(\ref{fceWeyl}),
and (\ref{tr-Weyl-brs}) in order to express them in coordinates
$\{ t$, $\r$, $z$, $\ffi\}$. As the~Weyl-Papapetrou metric
is stationary, via the~transformation (\ref{tr-Weyl-brs}), we get 
the~metric functions in the~stationary region bellow the~roof.
However, using the~same expressions for the~metric functions 
$\m$, $\l$, and $\o$ 
above the~roof, we obtain an analytical continuation of the~metric
(\ref{dsbrs}) across the~roof. It turns out that one has to change
the~sign of $\epsilon_3$ in order to find an analytical continuation
of the~metric across the~null cone of the~origin 
\BE
\r^2+z^2-t^2=R_3=0\ .\label{nullcone}
\EE

The~non-stationary region above the~roof was not included in 
the~Weyl-Papapetrou form, however, it was contained
in the~original metric (\ref{rotcmet}) in the~$\{ \t$, $x$, $y$, $\f\}$ 
coordinates.

\begin{figure}
\begin{center}
\includegraphics*{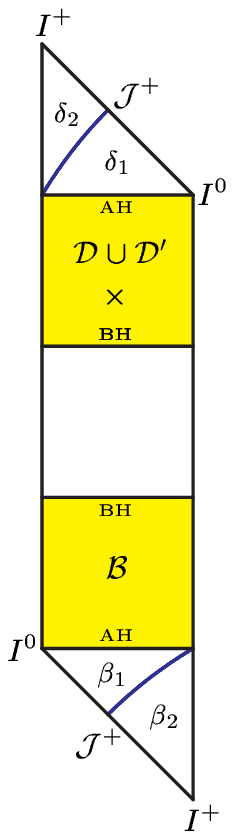}
\end{center}
\caption{This figure represents the~fourth column of Fig.~\ref{ctverce},
where points $y=\pm\infty$ are identified. The~diagram is cut along 
the~line $x+y=0$ where ${\cal J}^+$, $I^+$,  and $I^0$ are  located.
The~stationary  ${\cal B}$ and non-stationary $\b_1\cup\b_2$ regions
 correspond to the~regions I and II in Fig.~\ref{fig-boost}, respectively.
The~curves in both triangles represent the~null cones of the~origin
 (\ref{nullcone}) that divide them into the~regions
$\b_1$ and $\d_1$ outside
the~null cone and  $\b_2$ and $\d_2$ inside the~null cone.
The~stationary region ${\cal D}\cup {\cal D}'$ with the~curvature singularity
denoted by a cross and the~triangle above it represent the~inner part
of the~spacetime of a uniformly accelerated Kerr black hole, 
i.e., a uniformly accelerated
white hole with a ring singularity.
}
\label{fig-nekonecno}
\end{figure}

The~region above the~roof inside the~null cone of the~origin  
($F<0$ and $\epsilon_3>0$ in (\ref{eqRi2}), 
i.e., the~region $\b_2$ in Fig.~\ref{fig-nekonecno}) 
can be transformed
into the~cylindrically symmetric metric (see \cite{BicSchPRD} for 
the~case with hypersurface orthogonal Killing vectors)
\BE
{\rm d}s^2={\rm e}^{-2{\tilde U}}
        [ {\rm e}^{2{\tilde \n}} (-{\rm d}{\tilde  t}^2+{\rm d}{\tilde \r}^2)
      +{\tilde \r}^2{\rm d}{\tilde  \ffi}^2] 
+{\rm e}^{2{\tilde U}}({\rm d}{\tilde  z}
+{\tilde  \o}{\rm d}{\tilde  \ffi})^2\ , \label{ds-eros1}
\EE
where ${\tilde U}$, ${\tilde \n}$, 
and ${\tilde \o}$ are functions of ${\tilde \r}$
and ${\tilde  t}$, by the~transformation 
\BEA
{\tilde {\rho}}^2 &=&{{\cal K}}^2 \frac{-FG}{A^4 (x+y)^4}\ ,  
\nn \\ 
\tilde  t  &=& {{\cal K}} \frac{1+mAxy(x-y)+xy-a^2A^2x^2y^2}{A^2(x+y)^2}\ , 
\nn\\ 
\t &=& \kappa_1 {\tilde z}  + \kappa_2 {\tilde \ffi} \ ,
\nn\\ 
\f &=& \kappa_3 {\tilde z} + \kappa_4 {\tilde \ffi} \ ,\nn\\ 
{\cal K} &=& \kappa_2 \kappa_3 - \kappa_1 \kappa_4 \ .\nn 
\EEA
Then it can be transformed into the~boost-rotation symmetric metric 
(\ref{dsbrs}) by
\BEA
{\tilde \r}^2&=&A^2\r^2(t^2-z^2)\ ,\nn\\
{\tilde \ffi}&=&\ffi\ ,\nn\\ 
{\tilde t}-{\bar z}_3&=&\pul A(\r^2+t^2-z^2)\ ,\nn\\
{\tilde z}&=&\frac{1}{A}{\rm arctanh}\left(\frac{t}{z}\right)\ .\nn
\EEA

Similarly by the~transformation
\BEA
{\hat t}^2 &=& {\cal K}^2 \frac{-FG}{A^4 (x+y)^4}\ ,  
\nn\\ 
\hat {\rho} &=& {\cal K} \frac{1+mAxy(x-y)+xy-a^2A^2x^2y^2}{A^2(x+y)^2}\ , 
\nn\\ 
\t &=& \kappa_1 {\hat z}  + \kappa_2 {\hat  \ffi} \ ,
\nn\\ 
\f &=& \kappa_3 {\hat z} + \kappa_4 {\hat \ffi} \ ,\nn\\
{\cal K} &=& \kappa_2 \kappa_3 - \kappa_1 \kappa_4 \ ,
\nn
\EEA
the~region above the~roof outside the~null cone of the~origin   
($F<0$ and $\epsilon_3<0$ in (\ref{eqRi2}), 
i.e., the~region $\b_1$ in Fig.~\ref{fig-nekonecno}) 
can be transformed
into another cylindrically symmetric metric 
\BE
{\rm d}s^2={\rm e}^{-2{\hat U}}
        [ {\rm e}^{2{\hat \n}} (-{\rm d}{\hat t}^2+ {\rm d} {\hat \r}^2)
      +{\hat t}^2{\rm d}{\hat \ffi}^2] 
+{\rm e}^{2{\hat U}}({\rm d}{\hat z}+{\hat \o}{\rm d}{\hat \ffi})^2\ , 
\label{ds-eros2}
\EE
where the~metric functions ${\hat  U}$, ${\hat \n}$, 
and ${\hat \o}$ depend on ${\hat \r}$
and ${\hat t}$. 
Again, it can be 
transformed into the~form 
(\ref{dsbrs}) by
\BEA
{\hat t}^2&=&A^2\r^2(t^2-z^2)\ ,\nn\\
{\hat \ffi}&=&\ffi\ ,\nn\\
{\hat \r}-{\bar z}_3&=&\pul A(\r^2+t^2-z^2)\ ,\nn\\
{\hat z}&=&\frac{1}{A}{\rm arctanh}\left(\frac{t}{z}\right)\ .
\nn 
\EEA

Using inverse transformations from the~coordinates 
$\{ t$, $\r$, $z$, $\ffi\}$
into $\{ \t$, $x$, $y$, $\f\}$, 
we can find localization of $I^{0}$, $I^{+}$, and ${\cal J}^{+}$
as given in Fig.~\ref{fig-nekonecno}.

One can also determine the~${\cal J}^+$ location similarly as was done
in \cite{haw} for the~non-spinning C-metric. The~SC-metric can be compactified
by the~conformal factor $\Omega=A(x+y)$. The~future null infinity ${\cal J}^+$
is then at $\Omega=0$, i.e., $x+y=0$, with the~induced metric 
\BDM
{{\rm d}} {\tilde s}_{\cal J}^2=G(x)({\rm d} \t^2+{\rm d} \f^2)\ \nn
\EDM
and the~normal vector to ${\cal J}^+$
\BDM
{\bar n}={\tilde \nabla}^\a\Omega\ \der_{x^\a}
=\frac{AG(x)}{1+(aAx^2)^2}(\der_x-\der_y)\ .\nn
\EDM


\section{Geodesics in the spinning C-metric}
\label{sec-geod}

In this section, geodesics in the~stationary region {$\cal B$} of 
the~SC-metric, especially their stability, are examined.
As far as we know, the stability of circular orbits in stationary axisymmetric
spacetimes was only studied in the case with  an equatorial 
plane of symmetry\cite{bardeen,olda}. 
The~SC-metric (and also
other exact solutions, e.g., a superposition of two Schwarzschild or Kerr
black holes with different parameters) 
does not have an equatorial plane of symmetry 
(see Fig.~\ref{BaD-Weyl} a). 
In the~following, basic theorems on the~stability of solutions of ordinary
non-linear differential equations\cite{Walter} are employed.

Let us investigate the~stability of geodesics in a general stationary spacetime
with two Killing vectors, $\der /\der t$ and $\der /\der \f$. 
Later we will apply the~results obtained here to
the~SC-metric (\ref{rotcmet}). 
We assume  the~metric 
in coordinates $\{ x^0,\ x^1,\ x^2,\ x^3\}$$=\{ t,\ X,\ Y,\ \f\}$
to be of the~form
\BE
{\rm d}s^2=g_{XX}{\rm d}X^2+g_{YY}{\rm d}Y^2+g_{tt}{\rm d}t^2
+2g_{t\f}{\rm d}t{\rm d}\f+g_{\f\f}{\rm d}\f^2\ ,\label{dsstac}
\EE
where  all metric functions depend only on $X$ and $Y$
and $g_{XX}$ and $g_{YY}$ are positive.
Let us define ${\cal T}\rov  - g_{tt}$, ${\cal F} \rov g_{\f\f} $, 
and ${\cal W} \rov g_{t\f} $.

Since the~metric (\ref{dsstac}) has two Killing vectors, 
each freely falling particle
carries two conserved quantities
\BEA
U_\f&=&L =g_{\f\f}U^\f+g_{t\f}U^t={\cal F}U^\f+{\cal W}U^t\ ,\nn\\
U_t &=&-E=g_{t\f }U^\f+g_{tt }U^t={\cal W}U^\f-{\cal T}U^t\ ,\nn
\EEA
which imply
\BEA
\dot{\f}&\rov&\frac{{\rm d}\f}{{\rm d}\t}=U^\f
   =\frac{L{\cal T}-E{\cal W}}{{\cal F}{\cal T}+{\cal W}^2}\ ,\nn\\
\dot{t} &\rov&\frac{{\rm d}t }{{\rm d}\t}=U^t 
   =\frac{L{\cal W}+E{\cal F}}{{\cal F}{\cal T}+{\cal W}^2}\ .\label{UfLUtE}
\EEA

The~norm of the~four-velocity $U^\a=\dot{x}^\a$ then reads
\BE
-1 = g_{\a\b}U^\a U^\b
=g_{XX}\dot{X}^2 +g_{YY}\dot{Y}^2 
-{\cal R}-1\ 
\EE
with
\BE
{\cal R}\rov -1
-\frac{1}{{\cal F}{\cal T}+{\cal W}^2}
   (L^2{\cal T}-E^2{\cal F}-2EL{\cal W})\ .\label{R}
\EE

Thanks to (\ref{UfLUtE}), two geodesic equations are
satisfied identically  while the~other two read
\BE
 \ddot{X}+{\cal F}_1=0\ ,\mm 
  \ddot{Y}+{\cal F}_2=0\ ,\label{geodxy} 
\EE
where
\BEA
{\cal F}_1&=&\pul g^{XX}(\ \ g_{XX},_X \dot{X}^2+2g_{XX},_Y \dot{X}\dot{Y}
       -g_{YY},_X\dot{Y}^2-{\cal R},_X)\ ,\\
{\cal F}_2&=&\pul g^{YY}(-g_{XX},_Y \dot{X}^2+2g_{YY},_X \dot{X}\dot{Y}
       +g_{YY},_Y\dot{Y}^2-{\cal R},_Y)\ .
\EEA

The~system of two second-order differential equations (\ref{geodxy})
can be converted
to four first-order differential equations 
\BE
\dot{z}^\a={\cal F}^\a\ (z^\b)\ ,\mm
{\cal F}^\a=(z^2,\ -{\cal F}_1,\ z^4,\ -{\cal F}_2 )\label{soust}
\EE
introducing new variables $z^\a$
\BE 
z^1\rov X\ ,\mm z^2\rov \dot{X}\ ,\mm 
z^3\rov Y\ ,\mm z^4\rov \dot{Y}\ .
\EE

Stationary (equilibrium) points
of the system (\ref{soust}), $z^\a_0$, 
are given by ${\cal F}^\a(z^\b_0)=0$, i.e.,
\BEA
z^2&=&\dot{X}=0\ ,\mm z^4=\dot{Y}=0\ ,\mm 
\mbox{i.e.,}\mm {\cal R}(z_0^\a) =0\ ,\label{y_t}\\
{\cal F}_1&=&-\pul g^{XX}{\cal R},_X (z_0^\a) =0\ ,\label{F1}\\
{\cal F}_2&=&-\pul g^{YY}{\cal R},_Y (z_0^\a) =0\ .\label{F2}
\EEA
One can eliminate $E$ and $L$ from Eqs.~(\ref{F1}), (\ref{F2}) and derive
the~equation for the~stationary points $X_0$, $Y_0$
\BE
( {\cal F},_{[X}{\cal T},_{Y]})^2
-4({\cal T},_{[X}{\cal W},_{Y]})({\cal F},_{[X}{\cal W},_{Y]})=0  \ 
\label{x0y0}
\EE
and, using also Eq.~(\ref{y_t}), express $E$ and $L$ in terms 
of $X_0$ and $Y_0$.

Since the~system (\ref{soust}) is autonomous, its linearization
may help to determine the~stability of its stationary points\cite{Walter}.
The~linearized form of (\ref{soust}) in the~neighbourhood of 
 a stationary point $z_0^\a$ is
\BDM
\dot{z}^\a=A^\a_\b (z^\b-z_0^\b)\ ,\nn
\EDM
where 
\BDM
A^\a_\b\rov \frac{\der {\cal F}^\a}{\der {z^\b}} (z_0^\g) =
\begin{pmatrix}{0&1&0&0 \cr
       -{\cal F}^{(0)}_1,_X &0&-{\cal F}^{(0)}_1,_Y&0\cr
          0&0&0&1\cr
         -{\cal F}^{(0)}_2,_X&0&-{\cal F}^{(0)}_2,_Y&0}
\end{pmatrix}\nn
\EDM
is a constant matrix with eigenvalues
\BE
\l_\a= \pm \sqrt{\pul}\sqrt{-{\cal F}^{(0)}_1,_X-{\cal F}^{(0)}_2,_Y
        \pm\sqrt{({\cal F}^{(0)}_1,_X-{\cal F}^{(0)}_2,_Y)^2
+4{\cal F}^{(0)}_1,_Y{\cal F}^{(0)}_2,_X}}
\ ,\label{vlc}
\EE
with ${\cal F}^{(0)}_1,_X \rov {\cal F}_1,_X (z_0^\g)$, etc.
If all Re$\l_\a$  were negative the~equilibrium point
of (\ref{soust})
would be asymptotically stable and if at least one Re$\l_\a$  were positive
the~equilibrium point of (\ref{soust})
would be unstable.
If max$\{$Re$\l_\a\}$ were zero the~studied
point could be stable, however,
further examinations would be necessary. 
Since $\l_1+\l_2=\l_3+\l_4=0$, 
the~stationary point $z_0^\a$ is unstable if Re$\l_\a\not= 0$ for any $\a$
and could be stable only if  Re$\l_\a=0$ for all $\a$, i.e., for
\BEA
&&{\cal F}^{(0)}_1,_X{\cal F}^{(0)}_2,_Y>
{\cal F}^{(0)}_1,_Y{\cal F}^{(0)}_2,_X
\ ,\mm \mbox{i.e.,}\mm 
{\cal R}^{(0)},_{XX}{\cal R}^{(0)},_{YY}>
{\cal R}^{(0)},_{XY}\ ^2\ ,\nn\\
&&{\cal F}^{(0)}_1,_X> 0\ ,\mm{\cal F}^{(0)}_2,_Y> 0
\ ,\mm \mm\mm \ \mbox{i.e.,}\mm 
{\cal R}^{(0)},_{XX}< 0\ ,\mm{\cal R}^{(0)},_{YY}<0
\ ,\label{stabmax}
\EEA
i.e., if the~function ${\cal R}$ has a local
maximum in the~equilibrium point or if
\BEA
&&{\cal F}^{(0)}_1,_X={\cal F}^{(0)}_2,_Y=
{\cal F}^{(0)}_1,_Y{\cal F}^{(0)}_2,_X=0
\ ,\mm \nn\\
\mbox{i.e.,} &&
{\cal R}^{(0)},_{XX}={\cal R}^{(0)},_{YY}=
{\cal R}^{(0)},_{XY}=0\ .\label{stabmax0}
\EEA

In order to determine whether the~stationary points  given by
(\ref{x0y0}) and satisfying (\ref{stabmax}) or (\ref{stabmax0}) 
are indeed stable,  we employ the~Lyapunov method.
The Lyapunov function for (\ref{soust}) is a function $V$ of 
four independent variables $z^\a$ satisfying
in an open neighbourhood of an equilibrium point $z_0^\a$ 
the~conditions
\BEA
&&V(z^\a)> 0\mm \mbox{for}\mm z^\a\not=z_0^\a\ ,\mm V(z_0^\a)=0\ ,\label{V}\\
&&V'\rov
V,_{z^\a}{\cal F}^\a \leq 0 \ .\label{dotV}
\EEA
Notice that $V'$  is the~directional derivative of $V$
in the~direction of ${\cal F}^\a$, 
$V'=\lim_{t\msip 0}[V(z^\a+t{\cal F}^\a)-V(z^\a)]/t$.
If such a function exists then the~equilibrium point $z_0^\a$ is stable,
however, there is not a general method how to find  it.

It can be shown  that 
\BE
V=g_{XX}{z_2}^2+g_{YY}{z_4}^2- {\cal R}\ ,\label{LyapV}
\EE 
which satisfies $V'=0$ identically, 
is the~Lyapunov function for the~system (\ref{soust})
if ${\cal R}$ is negative in the~neighbourhood of the~equilibrium
point $z_0^\a$, i.e., if ${\cal R}$ has a local maximum
in $z_0^\a$ since ${\cal R}(z_0^\a)=0$. Thus, equilibrium points 
satisfying (\ref{stabmax}) are indeed stable.

In the~special case ${\cal W}=0$ with hypersurface orthogonal 
Killing vectors, 
${\cal R}\sim (E^2-V_{\mbox{ef}}^2)/E^2$,
where $V_{\mbox{ef}}$  is the~effective potential.
A local maximum of ${\cal R}$ then
corresponds to a local minimum of $V_{\mbox{ef}}$.

To summarize: The~stationary points $X=X_0$ and $Y=Y_0$
 given by 
(\ref{x0y0}) are stable if ${\cal R}$ (\ref{R})
has  a  local maximum there.

Let us apply the~foregoing general considerations
to the~SC-metric (\ref{rotcmet}).
The~corresponding computations 
are rather complicated and thus we present  only the~main results
here.

\begin{figure}
\begin{center}
\includegraphics*{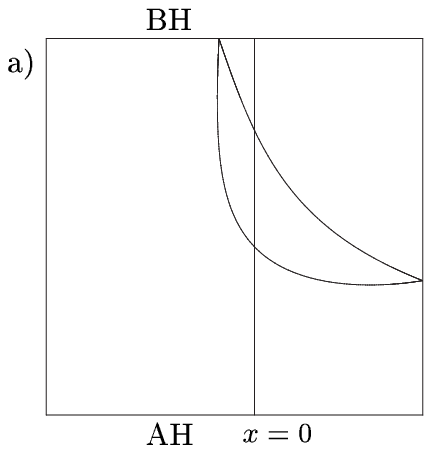} 
\hspace{1.5cm}
\includegraphics*{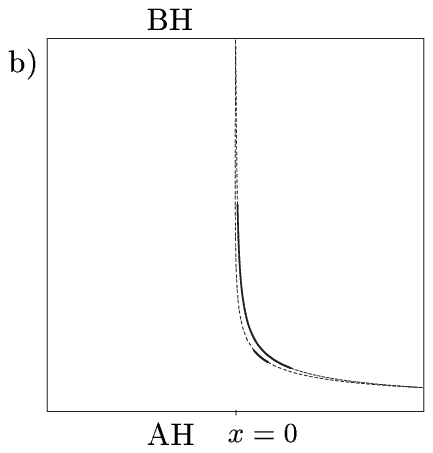}
\end{center}
\caption {Stationary points in the~${\cal B}$ region of the~SC-metric 
for a) $m=1$, $a=1/2$, $A=1/10$;
b) $m=1/5$, $a=1/6$, $A=1/100$ -- stable equilibrium points are
highlighted,
the~upper and lower curves correspond to retrograde and 
prograde orbits, respectively.}
\label{stacbody}
\end{figure}

The~condition (\ref{x0y0}) for stationary points
is a polynomial of the~order 12  in $x$ and $y$ and for
$a=0$ it reduces to Eq.~(10) in \cite{av-geod}.
The~stationary points $x_0$, $y_0$ in the~${\cal B}$ region 
are plotted in Fig.~\ref{stacbody}. As in the~non-spinning 
case\cite{av-geod}, the~stationary points $x_0=0$ correspond
to  null geodesics, stationary points with $x_0>0$ and $x_0<0$ 
to timelike and spacelike geodesics, respectively.

In contrast to the~non-spinning case, there now appear  two curves
in Fig.~\ref{stacbody} -- the~upper and  the~lower one corresponding 
to retrograde ($L<0$) and prograde ($L>0$) orbits, respectively.

Using Eq.~(\ref{stabmax}), one can show that for $A$ small,
there exist both retrograde and prograde stable orbits 
(Fig.~\ref{stacbody} b).
As was shown in \cite{av-geod}, these geodesics correspond
to co-accelerated  test particles orbiting the~uniformly 
accelerated black hole. If the~parameters $E$ and $L$ are perturbed
sufficiently the~test particle falls under the~black hole or acceleration
horizon.


\section{Conclusion}

If  conditions 
(\ref{podm-rootb}) are satisfied 
there are four stationary regions in the~SC-metric.
Each of them can be transformed into a different Weyl-Papapetrou metric by 
(\ref{transW1})--(\ref{transW5}). 
The~most physically important region ${\cal B}$ 
in the~Weyl-Papapetrou coordinates
represents the~gravitational field of a ``spinning rod'' 
(the~black hole horizon) and a semi-infinite line
mass that are held in equilibrium by a spinning string. There is 
an ergoregion in the~vicinity of the~black hole horizon
 and a region with closed
timelike curves in the~neighbourhood of the~spinning string (see Fig.~\ref{BaD-Weyl} a). 
In the~limit the~acceleration $A$ $\msip 0$, there remains just 
the~``spinning rod'', which corresponds to the~``exterior'' of the~Kerr metric. 
Through a further transformation to the~coordinates adapted
to the~boost and rotation symmetries new non-stationary radiative regions
appear. It turns out that these regions are already contained in
the~original form (\ref{rotcmet}) of the~SC-metric (see Fig.~\ref{fig-nekonecno}). 

Another stationary region ${\cal D}\cup{\cal D}'$ 
corresponds to a uniformly accelerated superposition
of a spinning white hole and a ring singularity, which 
represents a uniformly accelerated ``interior'' of
the~Kerr solution, i.e., the~region bellow
the~inner horizon up to $r\msip -\infty$. 
A timelike curve in the~Kerr metric can start in the~external,
asymptotically flat region, cross two horizons, 
pass through the~ring singularity, and 
emerge in another asymptotically flat white hole 
region, where gravity is repulsive.
Similarly, in the~SC-metric, a timelike curve 
starting in the~region ${\cal B}$ can cross
two horizons and debouch in the~region ${\cal D} \cup {\cal D}'$. 
Then, if it is not co-accelerated
it crosses the~null cone of the~origin and reaches 
$I^+$ (see Fig.~\ref{fig-nekonecno}).

In Sec.~\ref{sec-geod}, the~stability of geodesics corresponding to
equilibrium points in a general 
stationary spacetime with two symmetries was studied.
Using the~Lyapunov method  it can be shown that
an equilibrium point is stable if the~function ${\cal R}$ 
of two variables (\ref{R})
has a local maximum there. It turns out that for the~SC-metric
there exist stable prograde and retrograde orbits corresponding
to co-accelerated particles orbiting the~uniformly accelerated
spinning black hole.

\section*{Acknowledgments}

We are grateful to Professor J. {Bi\v c\' ak} for several
fruitful years of common work under his supervision, which also included 
the~SC-metric.
We  thank Professors I.~Babu\v ska and I.~Vrko\v c for several discussions
on the~stability of equilibrium points and 
M.~\v Z\' a\v cek for a discussion on the~papers \cite{bardeen,olda}.
We are also thankful to University of Texas, Austin, for
hospitality.

\end{document}